\documentclass[prl,twocolumn,aps,superscriptaddress]{revtex4} 

\usepackage{amsthm}
\usepackage{amsmath,latexsym,amssymb,verbatim,enumerate,graphicx}
\usepackage{hyperref}
\usepackage{color}
\usepackage{setspace}\usepackage{cancel}
\usepackage[colorinlistoftodos]{todonotes}
\usepackage[T1]{fontenc}
\usepackage[utf8]{inputenc}
\usepackage{mathtools}   
\usepackage{amsfonts}

\newtheorem{thm}{Theorem}
\newtheorem*{thm*}{Theorem}
\newtheorem{defi}[thm]{Definition} 
\newtheorem{prop}[thm]{Proposition}
\newtheorem*{prop*}{Proposition}

\newtheorem*{lem*}{Lemma}

\newtheorem*{fact*}{Fact}
\newtheorem{cor}[thm]{Corollary}
\newtheorem*{cor*}{Corollary}

\usepackage{prettyref}
\newcommand{\savehyperref}[2]{\texorpdfstring{\hyperref[#1]{#2}}{#2}}

\newrefformat{eq}{\savehyperref{#1}{\textup{(\ref*{#1})}}}
\newrefformat{lem}{\savehyperref{#1}{Lemma~\ref*{#1}}}
\newrefformat{def}{\savehyperref{#1}{Definition~\ref*{#1}}}
\newrefformat{thm}{\savehyperref{#1}{Theorem~\ref*{#1}}}
\newrefformat{cor}{\savehyperref{#1}{Corollary~\ref*{#1}}}
\newrefformat{cha}{\savehyperref{#1}{Chapter~\ref*{#1}}}
\newrefformat{sec}{\savehyperref{#1}{Section~\ref*{#1}}}
\newrefformat{app}{\savehyperref{#1}{Appendix~\ref*{#1}}}
\newrefformat{tab}{\savehyperref{#1}{Table~\ref*{#1}}}
\newrefformat{fig}{\savehyperref{#1}{Figure~\ref*{#1}}}
\newrefformat{hyp}{\savehyperref{#1}{Hypothesis~\ref*{#1}}}
\newrefformat{alg}{\savehyperref{#1}{Algorithm~\ref*{#1}}}
\newrefformat{rem}{\savehyperref{#1}{Remark~\ref*{#1}}}
\newrefformat{item}{\savehyperref{#1}{Item~\ref*{#1}}}
\newrefformat{step}{\savehyperref{#1}{step~\ref*{#1}}}
\newrefformat{conj}{\savehyperref{#1}{Conjecture~\ref*{#1}}}
\newrefformat{fact}{\savehyperref{#1}{Fact~\ref*{#1}}}
\newrefformat{prop}{\savehyperref{#1}{Proposition~\ref*{#1}}}
\newrefformat{prob}{\savehyperref{#1}{Problem~\ref*{#1}}}
\newrefformat{claim}{\savehyperref{#1}{Claim~\ref*{#1}}}
\newrefformat{relax}{\savehyperref{#1}{Relaxation~\ref*{#1}}}
\newrefformat{red}{\savehyperref{#1}{Reduction~\ref*{#1}}}
\newrefformat{part}{\savehyperref{#1}{Part~\ref*{#1}}}

\newcommand{\<}{\langle}
\renewcommand{\>}{\rangle}

\def\id{{\operatorname{id}}}

\DeclareMathOperator{\tr}{tr}
\def\be{\begin{equation}}
\def\ee{\end{equation}}
\def\ben{\begin{eqnarray}}
\def\een{\end{eqnarray}}
\def\ot{\otimes}

\def\bei{\begin{itemize}}
\def\eei{\end{itemize}}

\def\cE{{\cal E}}
\def\cG{{\cal G}}

\def\cP{{\cal P}}

\def\cS{{\cal S}}
\def\cO{{\cal O}}
\def\cH{{\cal H}}
\def\cL{{\cal L}}

\def\cS{{\cal S}}
\def\cH{{\cal H}}
\def\cL{{\cal L}}
\def\cB{{\cal B}}
\def\cA{{\cal A}}
\def\cR{{\cal R}}
\def\tr{{\rm Tr}}
\def\cF{{\cal F}}
\def\cA{{\cal A}}
\def\cB{{\cal B}}
\def\cC{{\cal C}}

\def\cV{{\cal V}}
\def\cU{{\cal U}}
\def\cK{{\cal K}}

\def\ctE{{\tilde{\cal E}}}
\def\ctF{{\tilde{\cal F}}}

\begin{document}
\title{Toy model of boundary states with spurious topological entanglement entropy}

\author{Kohtaro Kato}
\affiliation{Institute for Quantum Information and Matter  \\ California Institute of Technology, Pasadena, CA 91125, USA}

\author{Fernando G.S.L. Brand\~ao}
\affiliation{Institute for Quantum Information and Matter  \\ California Institute of Technology, Pasadena, CA 91125, USA}
\affiliation{Amazon Web Services, AWS Center for Quantum Computing, Pasadena, CA}

\begin{abstract}
Topological entanglement entropy has been extensively used as an indicator of topologically ordered phases. We study the conditions needed for two-dimensional topologically trivial states to exhibit spurious contributions that contaminates topological entanglement entropy. We show that if the state at the boundary of a subregion is a stabilizer state, then it has a non-zero spurious contribution to the region if and only if, the state is in a non-trivial one-dimensional $G_1\times G_2$ symmetry-protected-topological (SPT) phase. However, we provide a candidate of a boundary state that has a non-zero spurious contribution but does not belong to any such SPT phase. 
\end{abstract}
\maketitle

{\it Introduction.--}
Topologically ordered phases are gapped quantum phases that cannot be detected by conventional local order parameters. 
Topological entanglement entropy (TEE)~\cite{PhysRevLett.96.110404, PhysRevLett.96.110405} has been widely used as an indicator of such phases. 
For ground states in gapped two-dimensional (2D) models, the entanglement entropy $S(A):=-\tr\rho_A\log_2\rho_A$ of a region $A$ is expected to behave as 
\begin{equation}\label{eq:arealaw}
S(A)=\alpha|\partial A|-\gamma+o(1)\,
\end{equation}
where $\alpha$ is a constant, $\partial A$ is the boundary length, and $o(1)$ comprises terms vanishing in the limit of $|\partial A|\to \infty$. TEE is defined as the universal constant term $\gamma$~\cite{PhysRevLett.96.110404}. The term $\gamma$ is shown to be the logarithm of the total quantum dimension of the abstract anyon model under various conditions~\cite{PhysRevLett.96.110404,Kitaev2003a,PhysRevB.71.045110,BKI19}.

\begin{figure}[tbp]  
\begin{center}
\vspace{-2mm}
\includegraphics[width=8.6 cm]{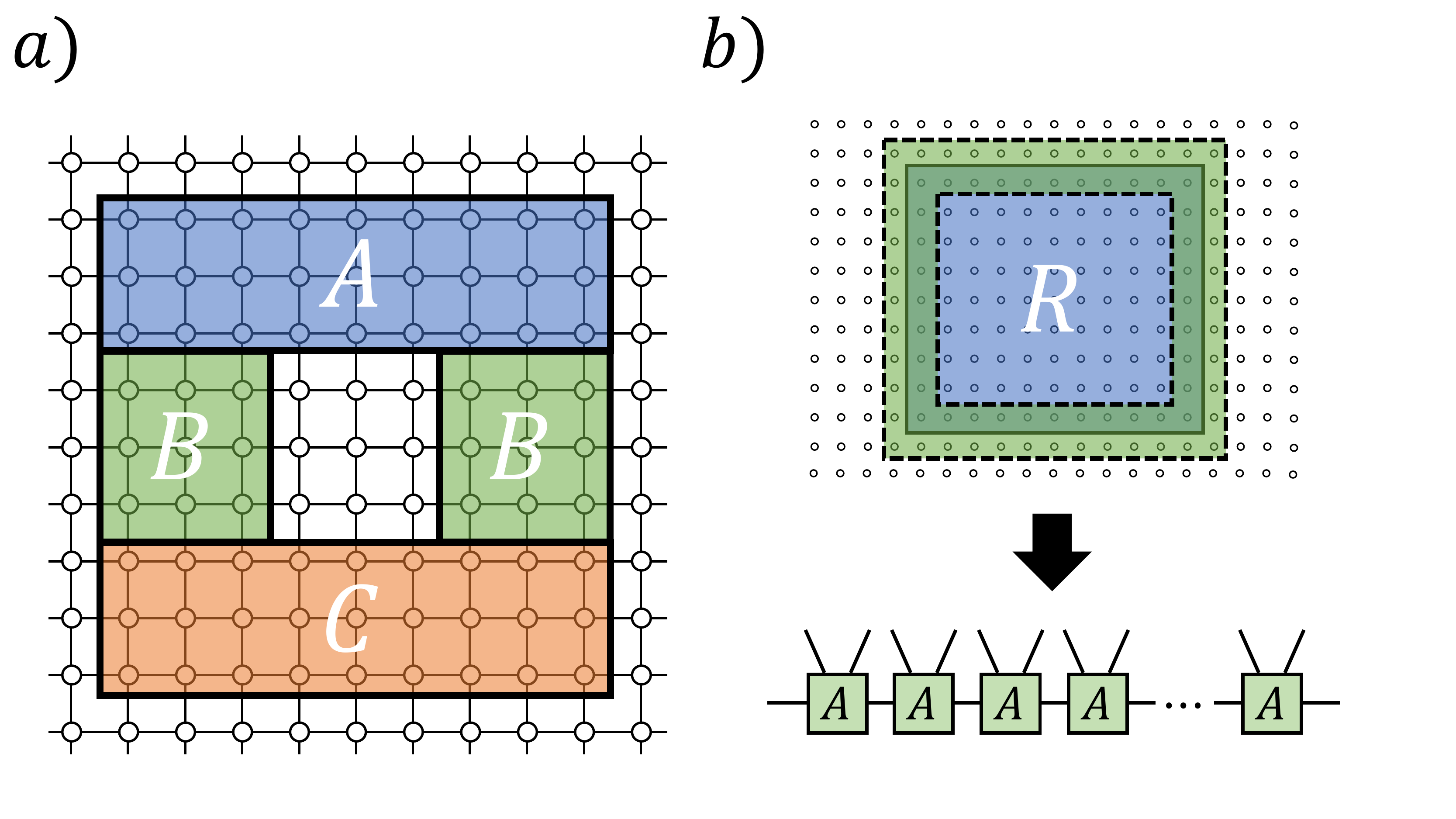}
\vspace{-7mm}
\end{center}
\caption{$a)$ A tripartition of a subsystem in a 2D spin lattice to calculate the TEE. $b)$ For ground states in the trivial phase, the entanglement entropy of region $R$ is determined by an MPS located at the boundary of $R$ (green region). Each tensor $A$ has two physical legs associated with $R$ and its complement respectively.
}\vspace{-4mm}
\label{fig:TEEmps}
\end{figure}

To extract the TEE from a ground state, one can calculate suitable linear combinations of entropies for certain subsystems (e.g., Fig.~\ref{fig:TEEmps}a), known as conditional mutual information (CMI) in quantum information theory, such that the first leading terms cancel out~\cite{PhysRevLett.96.110404, PhysRevLett.96.110405}. 

However, in general, Eq.~\eqref{eq:arealaw} could contain an additional term, and thus the above argument does not always work. This additional contribution, called spurious TEE~\cite{PhysRevB.94.075151,PhysRevLett.122.140506}, results in positive CMI for states in the trivial phase.

Thus far, the spurious TEE seems to be connected to the existence of a 1D symmetry-protected-topological (SPT) phase at the boundary of a certain region~\cite{PhysRevB.94.075151,PhysRevLett.122.140506,BravyiCEX,SSPT1,SSPT2,SSPT3}. Spurious TEE appears fragile against general local perturbations or small deformation of the regions, but the conditions under which spurious TEE appears are yet to be fully understood. 

A natural question is whether an SPT phase at the boundary is also a {\it necessary} condition for spurious TEE. 
In this letter, we study the underlying mechanism behind spurious TEE in the trivial phase. 
We model the degrees of freedom at the boundaries of regions (Fig.~\ref{fig:TEEmps}b) using matrix-product states (MPS)~\cite{MPSo08}.  
Here, we focus in particular on a renormalization fixed-point of the MPS in which the CMI is constant for all length scales. 
We then characterize the fixed points in terms of the operator-algebra quantum-error correction~\cite{OAQEC1,OAQEC2,OAQEC3}, and derive a formula to calculate the value of the spurious TEE from algebras associated to the single tensor. 

Using our characterization, we show that if the boundary MPS is a stabilizer state~\cite{stab}, a non-zero spurious TEE implies the MPS is in a non-trivial $G_1\times G_2$ SPT phase under on-site symmetry actions. By contrast, we also provide numerical evidence that, in general, there exist boundary states that have non-zero spurious TEE but do not belong to any such SPT phase. To the best of our knowledge, this is the first example of the mechanism of spurious TEE beyond the on-site $G_1\times G_2$ SPT phase at the boundary.

\vspace{0.2cm}
{\it MPS model of boundary states.-- }
We consider a translation-invariant ground state $|\psi\>$ defined on a 2D spin lattice of size $N$.  When a ground state is in the trivial phase, it can be (approximately) constructed from a product state only by a constant-depth local unitary circuit~\cite{PhysRevB.72.045141,PhysRevB.82.155138}.  
More precisely, there exists a set of unitaries $\{V_i\}$ such that 
\begin{equation*}\label{SRE}
|\psi\>=V_wV_{w-1}\ldots V_1|0\>^{\otimes N}\,,
\end{equation*}
where the depth $w=\cO(1)$ is a constant of $N$ and each $V_i$ is a product of local unitaries acting on disjoint sets of neighboring spins within radius $r=\cO(1)$.  

Let us divide the lattice into a connected region $R$ and its complement $R^c$. Entanglement between $R$ and $R^c$ is invariant under local unitaries $U_{R}U_{R^c}$; therefore, we can undo some parts of the circuit. Hence, $S(R)_\rho$ is equivalent to that of a tensor product of an entangled state $|\phi\>_{RR^c}$ around the boundary $\partial R$ and $|0\>$s elsewhere.   We call $|\phi\>_{RR^c}$ {\it the boundary state} of $R$ (Fig.~\ref{fig:TEEmps}b).

A constant-depth circuit can increase the Schmidt-rank by at most a constant. Therefore $|\phi\>_{RR^c}$ is written as an MPS (Fig.~\ref{fig:TEEmps}b): 
\begin{align*}
|\phi\>_{RR^c}=\sum&\tr(A^{i_1j_1}\ldots A^{i_{l}j_{l}})|i_1\ldots i_l\>_R|j_1\ldots j_l\>_{R^c}\,
\end{align*}
where $A^{i_kj_k}$ is a $D\times D$ matrix with a constant bond dimension $D=\cO(1)$.  
Here, we assume that all the tensors are the same due to the translation-invariance~\footnote{This assumption is slightly stronger than translation-invariance, since generally tensors can depend upon the direction of the edge and can even contain some ``corner'' tensors. However, such corner contributions cancel out in the calculation of CMI.}.  
Each local basis $\{|i_k\>\}$ corresponds to a coarse-grained site consisting of several neighboring spins so that the correlation length of the MPS is exactly zero.  We use $\cH$ and $\cK$ to denote the Hilbert spaces associated with $|i\>_R$ and $|j\>_{R^c}$, respectively. 
In this notation, there is an isometry $V:\mathbb{C}^D\ot\mathbb{C}^D\to\cH\ot\cK$, $V^\dagger V=I$, such that the MPS has the form~\cite{MPS1,MPS2}
\begin{align}\label{eq:pepsf}
|\phi\>_{RR^c}=V^{\ot l}|\lambda_D\>^{\ot l}\,
\end{align}
where $|\lambda_D\>=\sum_{k=1}^D\sqrt{\lambda_k}|kk\>$ is an entangled state with the Schmidt rank $D$. 
$V$ acts on two separated sites of neighboring $|\lambda\>$s.  
In the following, we especially consider the case in which $|\lambda_D\>$ is the maximally entangled state $|\omega_D\>:=\sum_{i=1}^D\frac{1}{\sqrt{D}}|ii\>$ for simplicity.  
We do not expect to lose much generality by this reduction, although we leave extension for future work. 

When $R$ is an annulus like $ABC$ in Fig.~\eqref{fig:TEEmps}a, we obtain two boundary states at the inner and outer boundaries.  
The ground state has a spurious TEE for $R$ if one of these boundary states has a non-trivial CMI
\begin{equation*}
I(A:C|B)_\rho:=S(AB)+S(BC)-S(B)-S(ABC)>0
\end{equation*}
for a tripartition $R=ABC$ such that $B$ separates $A$ from $C$.  
Importantly, the value of CMI matches that of the tri-information~\cite{PhysRevA.93.022317}, which is also used to extract TEE~\cite{PhysRevLett.96.110404} for the class of states that we are considering. 
 
\begin{figure}[tb]  
\begin{center}
\vspace{-2mm}
\includegraphics[width=8.6cm]{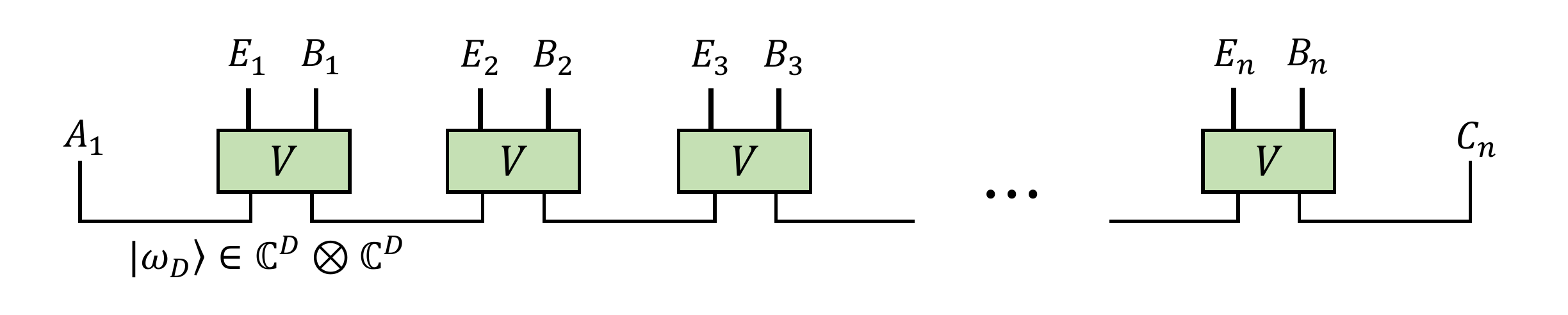}
\vspace{-7mm}
\end{center}
\caption{A schematic picture of the family of states $\phi^{(n)}$. $|\omega_D\>=\sum_i\frac{1}{\sqrt{D}}|ii\>$ is the $D$-dimensional maximally entangled state and $V$ is an isometry from $\mathbb{C}^D\ot\mathbb{C}^D$ to $\cH\ot\cK$. 
}
\label{fig:chain1}
\vspace{-3mm}
\end{figure}

Due to the monotonicity of CMI, a non-zero value of the spurious TEE implies that the CMI of an open-boundary MPS must be positive as well.  
We formalize a family of such open-boundary MPS $\{\phi^{(n)}\}_{n\geq0}$ with different lengths $n$ defined as
\begin{align*}
\phi^{(0)}&:=|\omega_D\>\<\omega_D|_{A_1A_2},\\
\phi^{(n)}&:=\cV_{A_{2n}A_{2n+1}\to B_nE_n}\left(\phi^{(n-1)}\ot|\omega_D\>\<\omega_D|_{A_{2n+1}A_{2n+2}}\right),
\end{align*}
where $\cV_{A_{2n}A_{2n+1}}(X)=VXV^\dagger$ is the isometry map (Fig.~\ref{fig:chain1}). For convenience, we relabel $A_{2n+2}$ by $C_{n}$ so that each $\phi^{(n)}$ is a state on $A_{1}\ot (B_{1}\ot E_1)\ot (B_{2}\ot E_2)\ot\cdots\ot (B_{n}\ot E_n)\ot C_{n}$. $A_1$ and $C_n$ represent the unfixed boundary condition.

After tracing out $R^c=E_1...E_n$, we have a family of mixed states
 $\{\rho^{(n)}\}_{n\geq0}$ defined by 
\begin{align*}
\rho^{(0)}&:=|\omega_D\>\<\omega_D|_{A_1A_2},\\
\rho^{(n)}&:=\cE_{A_{2n}A_{2n+1}\to B_n}\left(\rho^{(n-1)}\ot|\omega_D\>\<\omega_D|_{A_{2n+1}A_{2n+2}}\right),
\end{align*}
where $\cE=\tr_{E}\circ\cV$ is a completely positive and trace-preserving (CPTP) map. 
We denote the CPTP map $\cE(\cdot\ot|\omega_D\>\<\omega_D|)$ by $\ctE$. We then have 
\begin{align}\label{CPTPchain}
    \rho^{(n+1)}= \ctE_{C_{n}\to B_{n+1}C_{n+1}}(\rho^{(n)})\,.
\end{align}
The whole family is obtained by iteratively applying $\ctE$:
\begin{align}\label{eq:iter}
    \rho^{(n)}= \ctE_{C_{n}\to B_{n+1}C_{n+1}}\circ\cdots\circ  \ctE_{A_2\to B_1C_1}(\rho^{(0)})\,
\end{align}
 (recalling that $C_0=A_2$). We will simply denote the concatenated map in Eq.~\eqref{eq:iter} by $\ctE^{(n)}$. 
When we trace out $R$ instead of $R^c$, we obtain the complement chain, which we will denote by $\{\sigma^{(n)}\}$. We also define $\cF:=\tr_{B}\circ\cV$ and $\ctF(\cdot)=\cE^c(\cdot\ot|\omega_D\>\<\omega_D|)$.

$\{\phi^{(n)}\}$ has a spurious TEE if 
$I(A_1:C_n|B_1...B_n)_{\rho^{(n)}}$ is bounded from below by a positive constant.  
Although $\rho^{(n)}$ has zero correlation length, we might still have a non-trivial length scale for the CMI~\cite{Annals378}.  
We further remove this length scale by requiring saturation of the CMI:
\begin{align}\label{eq:satuCMI}
    I(A_1:C_1|B_1)_{\rho^{(1)}}=I(A_1:C_n|B_1\ldots B_n)_{\rho^{(n)}}\, \forall n.
\end{align}
Note that the LHS is always larger for any CPTP-map $\cE$. 
In the rest of this letter, we will simply denote $I(A_1:C_n|B_1B_2\ldots B_n)_{\rho^{(n)}}$ by $I(A_1:C_n|B_1B_2\ldots B_n)_{(n)}$. 

While the definition~\eqref{eq:satuCMI} depends on $n$, it is equivalent to two independent conditions independent of $n$. 
\begin{prop}\label{prop:satuCMIc} Eq.~\eqref{eq:satuCMI} is equivalent to 
\begin{align}
    I(A_1:B_1C_1)_{(1)}&=I(A_1:B_1B_2C_2)_{(2)}\,,\label{eq:saturateMI1}\\
    I(A_1:B_1)_{(1)}&=I(A_1:B_1B_2)_{(2)}\,.\label{eq:saturateMI2}
\end{align}
Moreover, Eq.~\eqref{eq:saturateMI2} is equivalent to 
\begin{align}\label{eq:saturateMI3}
        I(A_1:E_1C_1)_{(1)}&=I(A_1:E_1E_2C_2)_{(2)}\,.
\end{align} 
\end{prop}
Therefore, it is sufficient to consider up to $n=2$.

\vspace{0.2cm}
{\it Characterization by operator-algebra QEC.-- }
We use the theory of operator-algebra quantum-error correction (OAQEC)~\cite{OAQEC1,OAQEC2,OAQEC3} to characterize $\{\phi^{(n)}\}$. OAQEC is a general framework for quantum-error correction including standard codes~\cite{KL97} and subsystem codes~\cite{OQEC}. It allows us to describe what types of observables are correctable against a given error. For a given CPTP-map $\cE:\cH\to\cK$ representing a ``noise'', one can always specify the {\it correctable algebra} $\cA_\cE\subset\cB(\cH)$, which is a $C^*$-algebra containing all observables whose information is preserved under $\cE$ (see Supplemental Material (SM) for more details). 

In the following analysis, the correctable algebras of $\ctE$ and $\ctF$ play a crucial role. We first show that the saturation of the conditional mutual information~\eqref{eq:satuCMI} implies the saturation of these correctable algebras. 
\begin{prop}\label{prop:A=An}
If Eq.~\eqref{eq:satuCMI} holds for $\cE$, then
    \begin{align}
        \cA_{\ctE}&=\cA_{\ctE^{(n)}},\;\label{eq:p2eq1}\\ 
        \cA_{\ctF}&=\cA_{\ctF^{(n)}}\,,\\
           \cA_{\tr_C\circ\ctE}&=\cA_{\tr_C\circ\ctE^{(n)}},\; \\
\cA_{\tr_C\circ\ctF}&=\cA_{\tr_C\circ\ctF^{(n)}}\,,\forall n\,.\label{eq:p2eq2}
        \end{align}
\end{prop} 
This proposition means that the algebra $\cA_{\ctE}$ represents the information of the input, which is faithfully encoded in the output on $B_1...B_nC_n$ for all $n$. In the same way, $\cA_{\tr_C\circ\ctE}$ represents the perfectly recoverable information encoded on $B_1...B_n$. 

In general, there are operators carrying ``unpreserved'' information, which are disturbed and cannot be recovered perfectly. Such operators may also contribute to CMI, but may decrease it with respect to $n$. Prop.~\ref{prop:A=An} does not prevent such unpreserved operators and therefore the conditions~\eqref{eq:p2eq1}-\eqref{eq:p2eq2} are insufficient for Eq.~\eqref{eq:satuCMI}. In fact, we can always assume these conditions by coarse-graining a finite number of channels~\footnote{For any $\cE$, we have $\cB(\mathbb{C}^D)\supset\cA_{\ctE}\supset \cA_{\ctE^{(2)}} \supset\ldots\supset \mathbb{C}I$, and thus we cannot have an infinitely long sequence of strictly different $C^*$-algebras. Therefore, $m\in\mathbb{N}$ exists such that the conditions hold by redefining $\ctE\equiv\ctE^{(m)}$ ($\cF\equiv\ctF^{(m)}$). }. We utilize the concept of the complementary-recovery property~\cite{complementarity} because we intend to neglect this unpreserved information. 
\begin{defi}
We say that a CPTP-map $\cE$ satisfies the complementary-recovery property if 
\begin{equation}
    \cA_{\cE^c}=\cA_\cE'\;,
\end{equation}
where $\cA_\cE'$ is the commutant of $\cA_\cE$. 
\end{defi}
Any CPTP-map satisfies $\cA_{\cE^c}\subset \cA_\cE'$, i.e., any operator recoverable from the output of  the complementary channel $\cE^c$ should commute with the correctable algebra of the original channel $\cE$ (see also SM). As per the complementary-recovery property, the converse of this statement is also true. 
This property can be characterized by a projection map onto the correctable algebra. 
\begin{prop}
$\cE$ satisfies the complementary-recovery property if, and only if,
\begin{align}
    \cE\left(\cP_{\cA_\cE}(\rho)\right)&=\cE(\rho)\,,\forall \rho\,\\
    &or\nonumber\\
    \cP_{\cA_\cE}^\dagger\circ\cE^\dagger(O)&=\cE^\dagger(O)\,,\forall O\,,
\end{align}
where $\cP_{\cA_\cE}^\dagger:\cB(\cH)\to \cA_\cE$ is a conditional expectation onto $\cA_\cE$. 
\end{prop}
Therefore, the complementary-recovery property restricts the input information to the correctable algebra. 

\begin{defi}  We say isometry $V$ or CPTP-map $\cE$ satisfies dual complementarity if $\ctE$ and $\ctF$ both satisfy the complementary-recovery property.
\end{defi}
Dual complementarity reduces the four algebras in Prop.~\ref{prop:A=An}  to two algebras $\cA:=\cA_\ctE=(\cA_{Tr_C\circ\ctF})'$ and  $\cB:=\cA_\ctF=(\cA_{Tr_C\circ\ctE})'$. In what follows, we only consider states satisfying this property. 
We show that dual complementarity implies saturation of the CMI. Furthermore, its value is determined by $\cA$ and $\cB$. 
\begin{thm}\label{thm:converse}
If $V$ satisfies dual complementarity, then Eq.~\eqref{eq:satuCMI} holds. 
Let $\cA=\bigoplus_k M_{n_k}(\mathbb{C})\ot I_{n_k'}$ and $\cB=\bigoplus_lM_{m_l}(\mathbb{C})\ot I_{m_l'}$.  Then, the value of the CMI is given by
    \begin{align}\label{eq:cmiformula}
I(A_1:C_1|B_1)_{(1)}
&=\sum_kp_k\log \frac{n_k}{n_k'}+\sum_lq_l \log \frac{m_l}{m_l'}\,,
    \end{align}
where $p_k=\frac{n_kn_k'}{D}$ and $q_l=\frac{m_lm_l'}{D}$. Therefore, $I(A_1:C_1|B_1)_{(1)}>0$ if and only if
\begin{equation}\label{eq:strictincl}
\cB'\subsetneq\cA\,.
\end{equation}
\end{thm}
Eq.~\eqref{eq:strictincl} intuitively means an operator exists that is perfectly encoded in $BC$ whose information cannot be read out by just looking $B$.  Such non-local information causes a spurious contribution to CMI. 

Note that dual complementarity is not a necessary condition for Eq.~\ref{eq:satuCMI}. For example, one can consider $\cE=\cE'_{A_2}\ot \id_{A_3}$ such that $\cE'$ is not the completely depolarizing channel, but $\cA_{\cE'}=\mathbb{C}I$. The corresponding family satisfies $I(A:C|B)=0$ for any length, but the map does not satisfy dual complementarity. This is because all unpreserved information is transferred to $B$ and not $C$, and thus it cancels out in $I(A:C|B)=I(A:BC)-I(A:B)$.

\vspace{2mm}
{\it Relation to SPT phases.--} 
For any $O\in\cA$, we always find a corresponding {\it logical operator} ${\tilde O}_{BC}$ such that
\begin{align*}
 \cV_{A_2\to BEC} (O_{A_2}|\psi\>_{A_2})=  ({\tilde O}_{BC}\ot I_E)\cV_{A_2\to BEC}|\psi\>_{A_2}\
\end{align*}
for any $|\psi\>\in\mathbb{C}^D$, where 
\begin{align*}
    \cV_{A_2\to BEC}|\psi\>_{A_2}:=V^{\ot n}\left(|\psi\>_{A_2}\ot\bigotimes_i |\omega_D\>_{A_{2i+1}A_{2i+2}}\right)\,.
\end{align*} 
  
In general, the logical operator is not unique. The set of all logical operators $\cL_\cA$ in $\cA$ is given as the pre-image of $\ctE^{(n)\dagger}$: 
\begin{equation*}
\cL_\cA:=\{O_{BC}\, |\ctE^{(n)\dagger}(O_{BC})\in\cA\}\,.
\end{equation*}
 $\ctE^{(n)\dagger}$ is a normal $*$-homomorphism from the pre-image to $\cA$~\cite{OAQEC3}. By the first isomorphism theorem for algebra, the image of the homomorphism is isomorphic to the pre-image up to the kernel: 
\begin{equation*}
\cL_\cA/Ker\ctE^{(n)\dagger}\cong\cA\,. 
\end{equation*}
We denote the equivalence class of the logical operators of $O\in\cA$ by $\cL(O)$. 

Suppose that the boundary state is in a non-trivial SPT phase under a symmetry of group $G_1\times G_2$ acting on each tensor as $U(g_1,g_2)=U(g_1)_B\ot U'(g_2)_E$. The action induces a projective representation $V(g)\ot V(g)^\dagger$ on the virtual degrees of freedom~\cite{SPTMPS} (see also SM). For instance, it holds that
\begin{equation*}
U(g)_{B_1E_1}|\phi^{(1)}\>_{A_1B_1E_1C_1}=V(g)^T_{A_1}\ot V(g)_{C_1}^\dagger|\phi^{(1)}\>_{A_1B_1E_1C_1}
\end{equation*}
for $n=1$. This correspondence reads that $V(g_1)\in\cA$ and $V(g_2)\in\cB$. $V(g)$ has a logical unitary operator 
\begin{align*}\label{eq:tensorlogical}
    U(g)\ot U(g)\ot \cdots \ot U(g)\ot V(g)\in \cL(V(g))\, 
\end{align*}
whose support is $BC$ ($EC$) if $g=(g_1,e)$ ($g=(e,g_2)$). Suppose the state is in a non-trivial SPT phase in the sense that $[V(g_1),V(g_2)]\neq0$ for some $g_1,g_2$~\cite{PhysRevB.94.075151}. This implies that $\cB'\subsetneq\cA$. Therefore, we can reconfirm that the non-trivial $G_1\times G_2$ SPT phase implies non-zero CMI under dual complementarity.  

The converse direction is entirely non-trivial. The existence of tensor-product logical unitaries $U_B\ot U_C$ is necessary for $\phi^{(n)}$ to be a state in such an SPT phase, but this condition is not always implied by non-zero CMI, as we will observe later. 
A particular class of $V$ in which the converse also holds is isometry comprising Clifford gates, i.e., when the MPS is a stabilizer state~\cite{stab}. 
\begin{thm}\label{thm:spt}
 Let $V$ be an isometry composed of Clifford gates and ancillas $|0\>^{\ot k}$. Then
\begin{equation}
I(A_1:C_n|B_n)_{(n)}>0\ \forall n,
\end{equation}
if and only if finite groups $G_1$ and $G_2$ exist such that the MPS generated by $V$ is in a non-trivial $G_1\times G_2$ SPT phase. 
\end{thm}
The proof is given in SM. Theorem~\ref{thm:spt} can be applied for all 2D topologically trivial stabilizer states including the 2D cluster state~\cite{PhysRevLett.122.140506}. However, the conclusion is not necessarily true outside of stabilizer states. In fact, one can find a family of boundary states such that all non-identity logical unitaries cannot be written as $U_B\ot U_C$. 

\vspace{2mm}
{\it A non-trivial example.--}  
Let $V_{U}$ be an isometry that is the Stinespring dilation of $\cE_U(\sigma)=\frac{1}{4}\sum_{i=0}^3(P_i\ot P_iU)\sigma(P_i\ot U^\dagger P_i)$, where $P_i\;(i=0,1,2,3)$ are the Pauli Matrices ($P_0=I$). 
The correctable algebras are $\cA=\cB=M_2(\mathbb{C})$; therefore, boundary states automatically satisfy dual complementarity. The CMI attains a maximum value of $I(A:C|B)_{(n)}=2$. Note that 
this model is in a $D_2\times D_2$ SPT phase if $U=I$. 

\begin{figure}[b]  
\begin{center}
\vspace{-2mm}
\includegraphics[width=8.6cm]{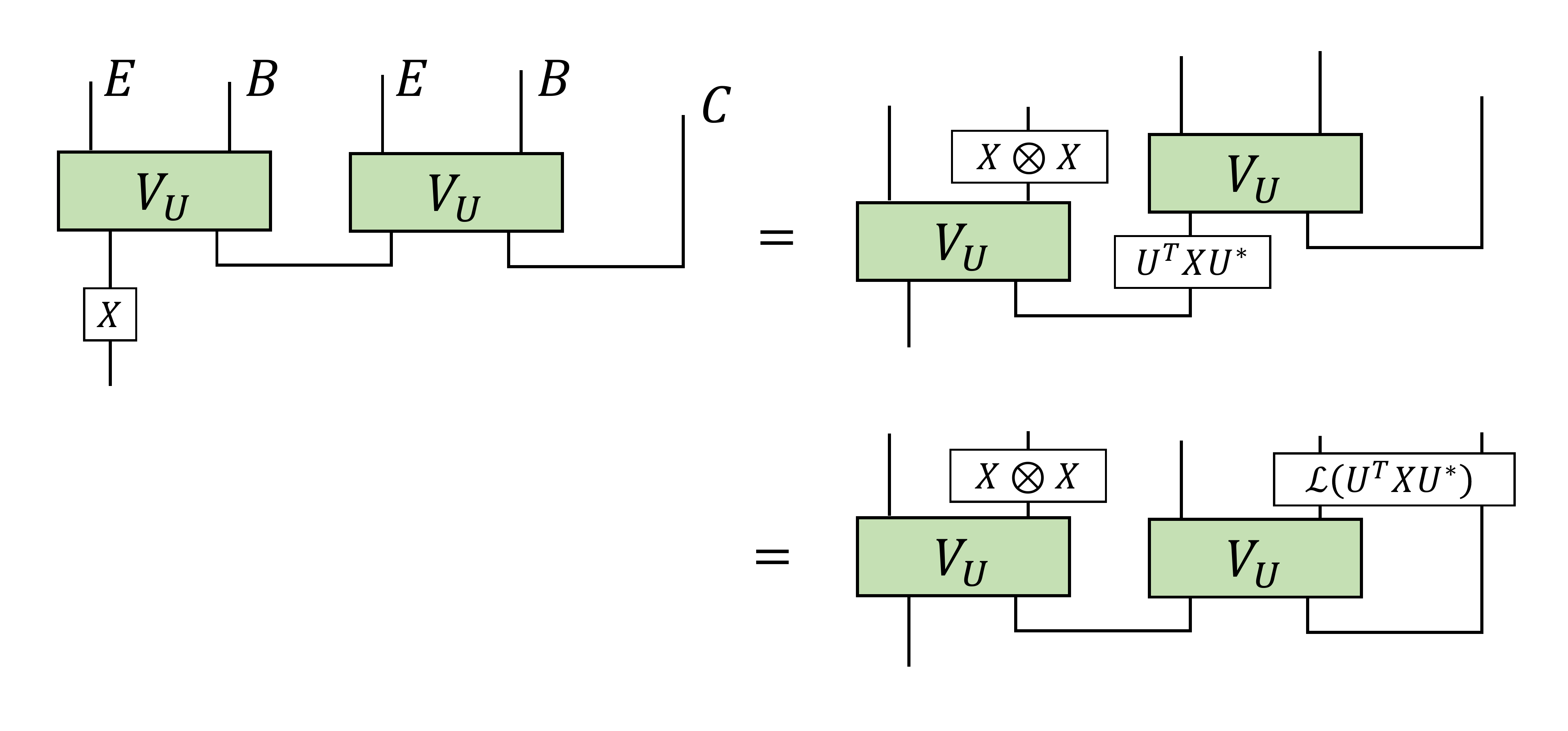}
\vspace{-8mm}
\end{center}
\caption{A logical operator of Pauli $X$ in the example. This is a unique logical operator of $X$ supported on $BC$. For general $U$, the logical operator is no longer a tensor product of unitaries on $B$ and $C$ for $n=2$. We expect this to hold for $n>2$ as well.}
\vspace{-0mm}
\label{fig:counter}
\end{figure}

For $n=1$, each Pauli operator $P_i$ has an unique logical operator $(P_iP_i)_B\ot U^T(P_i)_CU^*$~\footnote{The uniqueness follows from the fact that there is no stabilizer supported on $BC$.}. 
If $U$ is not a Clifford unitary nor diagonal in the $X$ or $Z$-bases, both $U^TX_CU^*$ and $U^TZ_CU^*$ are non-Pauli matrices. This induces non-tensor-product logical operators on $B_2C_2$, which are also unlikely to be a tensor product for $n>2$ (Fig.~\ref{fig:counter}). 
By coarse-graining $\ctE\equiv\ctE^{(2)}$, we obtain a model with no logical-operator form like $U_B\ot U_C$, but with $I(A:C|B)_{(n)}=2$.

We can construct a non-trivial 2D translation-invariant model by considering a layer of many copies of this 1D example along the vertical and the horizontal directions (decoupled stacks), as in the case of a 2D weak subsystem SPT phase~\cite{ssptorigin}. 
The resulting 2D state has a spurious TEE for an arbitrarily large dumbbell-like region~\cite{PhysRevLett.122.140506}. 

\begin{figure}[htbp]  
\begin{center}
\vspace{2mm}
\includegraphics[width=8.6cm]{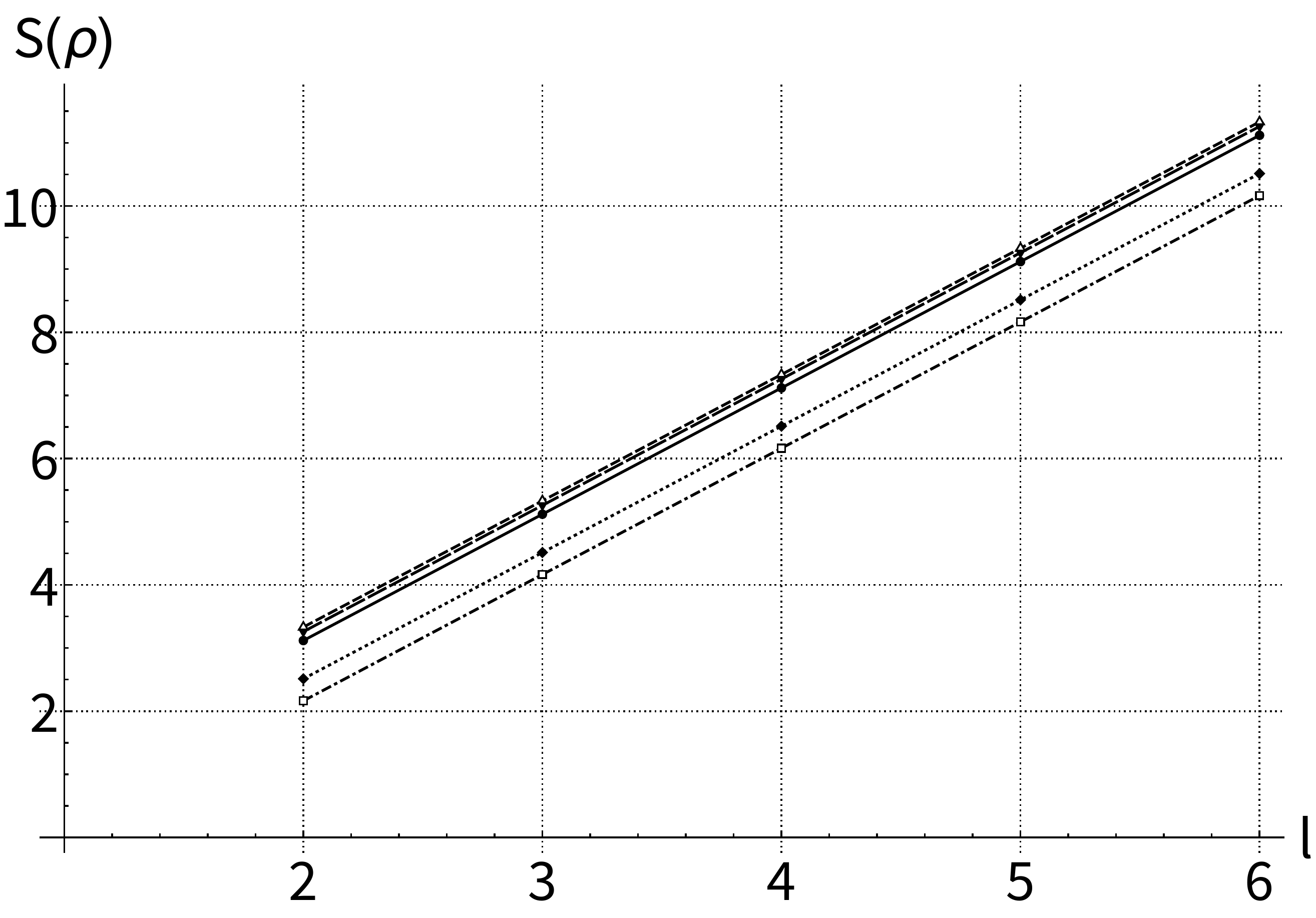}
\vspace{-8mm}
\end{center}
\caption{The numerical result on the entropy $S(\rho_{ABC})$ of the example with a closed boundary for 5 samples of $U$ from the Haar measure. $l$ is the length of the spin chain ($l=6$ is 12-qubit). From the data, $S(\rho)=2l-c_0$ with a constant $c_o>0$ up to $10^{-7}$ error. Since any reduced state of the example is completely mixed, it shows that $I(A:C|B)_{(n)}=c_0$ for any tripartition $ABC$ such that $B$ separates $A$ from $C$.}
\vspace{-3mm}
\label{fig:graph}
\end{figure}

One may expect that, for a periodic boundary condition, the CMI could vanish in such a non-trivial example. Although we do not have any analytical result for this, we numerically sampled $U$ from the Haar measure and then calculated the CMI for closed chains. Fig.~\ref{fig:graph} suggests that CMI remains a positive constant even for the closed boundary, while the value decreases from 2.

{\it Future directions.--} 
A crucial open question is how to characterize/classify the non-trivial example with spurious TEE. Although it should not be in a SPT phase under the on-site $G_1\times G_2$ symmetry, it could be in a SPT phase under another type of symmetry. 

Generalization to more broad classes of boundary MPS is desired. One possible extension is to consider boundary states without dual complementarity. Dual complementarity neglects all information outside of $\cA$, but in general, one has some ``noisy'' information localized on $B$ (or $E$). Extending the correctable algebra by adding operators carrying such information may be possible. 
Another important direction is considering general injective MPS including~\eqref{eq:pepsf}. We expect that general injective MPS can be decomposed into protected and unprotected parts, as in Ref.~\cite{SSPT3}, such that the effect of the unprotected part vanishes exponentially as the conditioning system grows. We leave these problems for future works. 

The spurious contribution exsists not only in the trivial phase but also in  topologically ordered phases. Although the boundary MPS construction does not work in topologically ordered phases, we can construct a non-trivial example of topologically ordered models with spurious contribution by taking tensor product of any 2D topological order model and the models considered in this paper. Such a state is could be considered as a symmetry-enriched topologically ordered phases beyond on-site symmetry.

{\it Acknowledgments.--} KK thanks Arpit Dua for a discussion about stacking 1D models to construct a 2D model.
KK acknowledge funding provided by the Institute for Quantum Information and
Matter, an NSF Physics Frontiers Center (NSF Grant {PHY}-{1733907}).
FB is supported by the NSF. The numerical calculation in this paper is implemented in Caltech HPC center.

\onecolumngrid
\section{Operator Algebra Quantum Error Correction}\label{ap1}
In this section, we summarize the theory of operator-algebra quantum error correction (OAQEC)~\cite{OAQEC1a}. 
Formally, we say that a set $\cS\subset\cB(\cH)$ on a code subspace $\cH_{C}=P_{C}\cH$ ($P_C=I$ in the main text) is correctable for $\cE$ if there exists a CPTP-map (also called a channel) $\cR$ such that 
\begin{align}
P_C\left(\cR\circ\cE\right)^\dagger(X)P_C=P_CXP_C\,,\forall X\in \cS\,.
\end{align}
We will call $\cR$ {\it recovery map}. 
This is equivalent to say we require the information of not the whole code space but particular observables are preserved: 
\begin{align}
\tr(X\rho)=\tr\left(X(\cR\circ\cE)(\rho)\right)\;\forall X\in S, \,\forall \rho\in\cS(\cH_C)\,.
\end{align}
This is in contrast to the standard (or subsystem) QEC in which we require that all information (= full density matrix, i.e. the expectation value for all observables) of the logical state is recoverable. A main result of Ref.~\cite{OAQEC1a} is the following:
\begin{prop}~\cite{OAQEC1a} Let $\cA$ be a subalgebra of $\cB(P_C\cH)$. Then, $\cA$ is correctable for $\cE$ on $\cH_C$ if, and only if,
\begin{align}
[P_CE_a^\dagger E_bP_C, X]=0 \;\forall X\in \cA, \forall a,b\,
\end{align}
i.e. the algebra $Alg(\{P_CE_a^\dagger E_bP_C\}_{a,b})$ is a subalgebra of $\cA'$, where $\cA'$ is the commutant of $\cA$. 
\end{prop}
In other words, any subalgebra of $Alg(\{P_{C}E_{a}^{\dagger}E_{b}P_{C}\}_{a,b})' $ is correctable. For a given code subspace $\cH_{C}$ and an error $\cE$, we define the {\it correctable algebra} of $\cE$ by
\begin{align}
\cA_\cE&:=Alg(\{P_CE_a^\dagger E_bP_C\}_{a,b})'\\
&=\left\{ X\in \cB(\cH_C)\;|\;[P_CE_a^\dagger E_bP_C,X]=0\;\forall a,b \right\}\,.
\end{align}
This is a unital $C^*$-subalgebra of $\cB(\cH_C)$. 

\subsection{A complementary relation for OAQEC}
In the standard QEC, we have a complementary relation implied by the no-cloning theorem: if one can exactly recover a logical state from a noise channel, then one cannot learn any information about the logical state from the output, i.e. the complimentary channel of the noise destroys all information about the logical state. In OAQEC, we can require to recover only partial information from noise. Hence, it might be possible to learn something from the outputs of the complementary channel. 

Suppose $\{E_{a}\}$ are Kraus operators of $\cE$. Then, we have a Steinespring isometry $V=\sum_{a}E_{a}\ot|\phi_{a}\rangle$ for an orthonormal basis $\{|\phi_{a}\rangle\}$ on $\cH_{E}$. The complementary channel $\cE^{c}$  has a Kraus representation by $\{F_{a}\}$, defined as
\begin{align}
F_a=\sum_b|\phi_b\>\<a|E_b\,,
\end{align}
where $\{|a\>\}$ is an orthonormal basis of $\cH_B$. The correctable algebra of $\cE^{c}$ is spanned by operators commuting with all
\begin{align}
P_CF_a^\dagger F_bP_C=P_C\cE^\dagger(|a\>\<b|)P_C
\end{align}
which span $P_{C}(Im\cE^{\dagger})P_{C}$. Hence, $\cA_{\cE^{c}}=Alg[(P_{C}(Im\cE^{\dagger})P_{C})]'$. 

Any $X\in\cA_{\cE}$ satisfies $X=P_{C}\cE^{\dagger}\circ\cR^{\dagger}(X)P_{C}$, and therefore $\cA_{\cE}\subset\cA_{\cE^{c}}'$: the correctable algebra of a channel is in the commutant of the correctable algebra of the complementary channel. Information of operators in $\cA_{\cE}\cap\cA_{\cE^{c}}$ can be extracted from the outputs of both channels and must be in the centers of $\cA_{\cE}$ and $\cA_{\cE^{c}}$. For example, if $\cA_{\cE}=\cB(\cH_{C})$, then $\cA_{\cE^{c}}\subset\cB(\cH_{C})^{'}=\mathbb{\ensuremath{C}}I$ which corresponds to the complementary relation for the standard QEC. 

\section{MPS classification of $G_1\times G_2$ SPT phases}
Here we briefly summarize the MPS classification of 1D SPT phases under on-site  $G=G_1\times G_2$ symmetry~\cite{SPTMPSa,PhysRevB.94.075151a} to be self-contained. Consider a MPS defined as
\begin{equation}
|\psi_N\>=\sum_{{\bf i},{\bf j}}\tr\left(A^{i_1j_1}A^{i_2j_2}\ldots A^{i_Nj_N}\right)|i_1i_2\ldots i_N\>_{R}|j_1j_2\ldots j_N\>_{R}\,,
\end{equation}
where $i_k,j_k=1,\ldots,d,\; \forall k$ and $A^{ij}$ is a $D\times D$ matrix. Consider a finite group  $G_1\times G_2$ with a unitary representation $(g_1,g_2)\mapsto U(g_1)\ot U'(g_2)$ acting on $\mathbb{C}^d\ot\mathbb{C}^d$. $G_1$ ($G_2$) then acts on $R$ $(R^c)$ as an on-site global symmetry $U(g_1)^{\ot N} \left(U'(g_2)^{\ot N}\right)$. Suppose $|\psi_N\>$ is a symmetric state 
\begin{equation}
\left(U(g_1)\ot U'(g_2)\right)^{\ot N}|\psi_N\>=|\psi_N\>\,,\quad \forall (g_1,g_2)\in G_1\times G_2\,
\end{equation}
for all $N$ (up to a global phase). This condition is shown to be equivalent to the following conditions for the single site:
\begin{equation}
\sum_{k}U(g_1)_{ik}A^{kj}=e^{i\theta(g_1)}V(g_1)A^{ij}V(g_1)^\dagger\,,\quad \sum_{k}U'(g_2)_{jk}A^{ik}=e^{i\theta(g_2)}V(g_2)A^{ij}V(g_2)^\dagger\,.
\end{equation}
Here, $V(g_1)$ and $V(g_2)$ are unitary consisting a projective representation of $G_1\times G_2$ in together. 

1D SPT phase under symmetry $G$ is classified by the second group cohomology $H^2(G;U(1))$ which is a group of the equivalence class $[\Theta]$ of the phase factor $\Theta:G\times G\to U(1)$. 
Each factor $\Theta$ is defined from the projective representation via
\begin{equation}
V(gh)=e^{i\Theta(g,h)}V(g)V(h)\,
\end{equation}
up to the equivalence relation
\begin{equation}
\Theta(g,h)\sim \Theta(g,h)+\theta(gh)-\theta(g)-\theta(h) \quad{\rm mod}\; 2\pi\,.
\end{equation}
The symmetric state $|\psi_N\>$ is said to be in a non-trivial $G$ SPT phase if the projective representation $V(g)$ is associated to a non-trivial element in the second cohomology class. In particular, in this paper, we employ the following definition used in Ref.~\cite{PhysRevB.94.075151a}:
\begin{defi}
We say symmetric MPS $|\psi_N\>$ is in a non-trivial $G_1\times G_2$ SPT phase if there is $(g_1,g_2)\in G_1\times G_2$ such that
\begin{equation}
[V(g_1), V(g_2)]\neq0\,.
\end{equation}
\end{defi}
This only happens for non-trivial projective representation since $[U(g_1),U'(g_2)]=0$. 

\section{Proofs of the theorems}
\subsection{Proof of Proposition 1}
{\it proof.} For any CPTP-map $\cE$, the corresponding family $\{\rho^{(n)}\}$ satisfies
\begin{align*}
    I(A_1:C_1|B_1)_{(1)}&= I(A_1:B_1C_1)_{(1)}-I(A_1:B_1)_{(1)}\\
    &\geq I(A_1:B_1B_2C_2)_{(2)}-I(A_1:B_1)_{(1)}\\
    &=I(A_1:B_1B_2C_2)_{(2)}-I(A_1:B_1)_{(2)}\\
    &=I(A_1:C_2|B_1B_2)_{(2)}+I(A_1:B_2|B_1)_{(2)}\\
    &\geq I(A_1:C_2|B_1B_2)_{(2)}\,.
\end{align*}
The first inequality follows from the data-processing inequality~\cite{DPI}  and Eq.~(3), and the second inequality follows from the strong subadditivity~\cite{SSA}. Equality holds when the both inequalities are saturated.  The first inequality is saturated if, and only if,~\cite{DPI}: 
\begin{equation}
    I(A_1:B_1C_1)_{(1)}=I(A_1:B_1B_2C_2)_{(2)}\,.
\end{equation}
The second inequality saturated if and only if $I(A_1:B_2|B_1)_{(2)}=0$, i.e. 
\begin{align*}
    S_{A_1}^{(1)}+S_{B_1}^{(1)}-S_{A_1B_1}^{(1)}=S_{A_1}^{(2)}+S_{B_1B_2}^{(2)}-S_{A_1B_1B_2}^{(2)}\,,
\end{align*}
which is equivalent to Eq.~(7) by $\rho^{(1)}_{A_1B_1}=\rho^{(2)}_{A_1B_1}$. 
Since $\rho^{(1)}_{A_1B_1E_1C_1}$ and $\rho^{(2)}_{A_1B_1B_2E_1E_2C_2}$ are pure states, the above equality is equivalent to 
\begin{align*}
    -S_{A_1}^{(1)}+S_{A_1E_1C_1}^{(1)}-S_{E_1C_1}^{(1)}=-S_{A_1}^{(2)}+S_{A_1E_1E_2C_2}^{(2)}-S_{E_1E_2C_2}^{(2)}\,
\end{align*}
after subtracting $2S_{A_1}^{(1)}=2S_{A_1}^{(2)}$, which is Eq.~(8). 
Therefore, $I(A_1:C_1|B_1)_{(1)}=I(A_1:C_2|B_1B_2)_{(2)}$ if and only if  Eqs.~(6,8) holds. The same argument applies to $I(A_1:C_1|E_1)_{(1)}=I(A_1:C_2|E_1E_2)_{(2)}$. 

We now show Eqs.~(6,8) imply
\begin{align*}
    I(A_1:B_1C_1)_{(1)}&=I(A_1:B_1\ldots B_nC_n)_{(n)}\,, \forall n\,,\\
    I(A_1:E_1C_1)_{(1)}&=I(A_1:E_1\ldots E_nC_n)_{(n)}\,, \forall n\,.
\end{align*}
The converse is clear. 
Since $\rho^{(2)}=\ctE_{C_1\to B_2C_2}(\rho^{(1)})$, Eq.~(6) implies there exists a \emph{recovery map} $\hat{\cR}_{B_1B_2C_2\to C_1}$ such that 
\begin{align}
    \hat{\cR}_{B_1B_2C_2\to C_1}(\rho^{(2)})&=\hat{\cR}_{B_1B_2C_2\to C_1}\circ\ctE_{C_1\to B_2C_2}(\rho^{(1)})\\
    &=\rho^{(1)}\,.\label{eq:rr=r}
\end{align}
By definition, we can write $\rho^{(3)}$ as $\ctE^L_{A_3\to A_1B_1}(\rho^{(2)}_{A_3B_2B_3C_3})$, where $\ctE^L_{A_3\to A_1B_1}(\cdot):=\cE_{A_2A_3\to  B_1}(\omega_{A_1A_2}\ot\cdot)$ is a CPTP-map (here we used the notation $\rho_{A_3B_2B_3C_3}^{(2)}$ for the state after shifting $\rho_{A_1B_1B_2C_2}^{(2)}$). From this expression, Eq.~\eqref{eq:rr=r} implies 
\begin{align}
    \hat{\cR}_{B_2B_3C_3\to C_2}(\rho^{(3)})&=
    \hat{\cR}_{B_2B_3C_3\to C_2}\circ\ctE_{C_2\to B_3C_3}(\rho^{(2)})\\
    &=\ctE^L\circ\hat{\cR}_{B_2B_3C_3\to C_2}\circ\ctE_{C_2\to B_3C_3}(\rho^{(1)})\\
    &=\ctE^L(\rho^{(1)})\\
    &=\rho^{(2)}\,.
\end{align}
This is equivalent to 
\begin{align}
    I(A_1:B_1B_2C_2)_{(2)}=I(A_1:B_1B_2B_3C_3)_{(3)}\,.
\end{align}
The same argument works for $n>3$ and under exchanging of $B\leftrightarrow E$.  Therefore Eqs. (6,8) are equivalent to Eq.~(5), which completes the proof. 
\qed

\subsection{Proof of Proposition~2}
\noindent {\it Proof.} By definition, it is clear that $\cA_{\ctE}\supset\cA_{\ctE^{(n)}}$ (information which is recoverable after an additional noise is recoverable without the noise). Suppose that Eq.~(5) holds. Then, as shown in the proof of Prop.~1, there exists a recovery map associated with the data-processing inequality such that
\begin{align}
    \hat\cR_{B_1B_2C_2\to B_1C_1}\circ\ctE^{(2)}_{C_0\to B_1B_2C_2}=\ctE^{(1)}_{C_0\to B_1C_1}\,.
\end{align}
There also exists another recovery map $\cR$ associated with OAQEC, such that
\begin{align}
    (\ctE^{(1)}_{C_0\to B_1C_1})^\dagger\circ\cR_{B_1C_1\to C_0}^\dagger(X_{C_0})=X_{C_0}
    \end{align}
for any $X_{C_0}\in \cA_{\ctE}$. By combining these two recovery maps together, we obtain that 
\begin{align}
    (\ctE^{(1)}_{C_0\to B_1C_1})^\dagger\circ\cR_{B_1C_1\to C_0}^\dagger(X_{C_0})&=(\ctE^{(2)}_{C_0\to B_1B_2C_2})^\dagger\circ\hat{\cR}^\dagger_{B_1B_2C_2\to B_1C_1}\circ\cR_{B_1C_1\to C_0}^\dagger(X_{C_0})\nonumber\\
    &=X_{C_0}\,.
\end{align}
Hence $\cR\circ\hat{\cR}$ can be regarded as the recovery map (in the sense of OAQEC) against the noise $\ctE^{(2)}$. Therefore, any $X_{C_0}\in \cA_{\ctE}$ is recoverable after applying $\ctE^{(2)}$ and thus $X_{C_0}\in \cA_{\ctE^{(2)}}$. By repeating the same  argument for all $n$ and the complementary channel, we complete the proof.  
\qed

\subsection{Proof of Proposition~4}
\noindent {\it Proof.} It is always true that $\cA_{\cE^c}=Alg(Im\cE^\dagger)'$. If $\cE$ satisfies the complementary recovery, $\cA_\cE=\cA_{\cE^c}'=Alg(Im\cE^\dagger)$. Therefore, $Im\cE^\dagger \subset Im\cP^\dagger_{\cA_\cE}$ and Eq.~(15) holds. Conversely, if $Im\cE^\dagger \subset Im\cP^\dagger_{\cA_\cE}$, then $Alg(Im\cE^\dagger)\subset \cA_\cE$ and therefore $\cA_{\cE^c}'\subset \cA_\cE$. This completes the proof since $\cA_\cE \subset \cA_{\cE^c}'$ always holds.  
\qed

\subsection{Proof of Theorem~6}
\noindent {\it Proof.} Recall that we have shown that $I(A_1:B_1C_1)_{(1)}=I(A_1:B_1B_2C_2)_{(2)}$ (and that of $E$) is sufficient for Eq.~(5).  $\cA_{\ctE}=\cA_{\ctE^{(2)}}$ (we assume this w.l.o.g.) implies that there exists $\cR$ such that $\cR\circ\ctE^{(2)}$ act as the identity map on $\cA_{\ctE}$. It holds that  $\ctE\circ\cR\circ\ctE^{(2)}=\ctE\circ\cP_{\cA_\ctE}\circ\cR\circ\ctE^{(2)}=\ctE$, so we can set $\hat{\cR}:=\ctE\circ\cR$. This completes the proof by the data-processing inequality.  

To calculate CMI, let us show  $I(A_1:B_1C_1)_{(1)}=I(A_1:A_2)_{\omega^\cA_D}$. Since $\ctE\circ\cP_\cA=\ctE$ by dual complementarity, we have $\ctE_{A_2\to B_1C_1}(\omega^\cA_{D A_1A_2})=\rho^{(1)}_{A_1B_1C_1}$. By the monotonicity of the mutual information under local CPTP-maps, this implies $I(A_1:B_1C_1)_{(1)}\leq I(A_1:A_2)_{\omega^\cA_D}$. By applying the recovery map of OAQEC $\cR_{B_1C_1\to A_2}$, we also have $\cP_{\cA}\circ\cR_{B_1C_1\to A_2}(\rho^{(1)}_{A_1B_1C_1})=\omega_D^\cA$, therefore $I(A_1:B_1C_1)_{(1)}\geq I(A_1:A_2)_{\omega^\cA_D}$. The same arguments hold for the complementary channel and $\cB$, and thus we have 
\begin{align}
I(A_1:C_1|B_1)_{(1)}&=I(A_1:B_1C_1)_{(1)}-I(A_1:B_1)_{(1)}\\
&=I(A_1:B_1C_1)_{(1)}+I(A_1:E_1C_1)_{(1)}-2S(A_1)\\
&=I(A_1:A_2)_{\omega^\cA_D}+I(A_1:A_2)_{\omega^\cB_D}-2S(A_1)\\
&=I_c(A_1\>A_2)_{\omega^\cA_D}+I_c(A_1\>A_2)_{\omega^\cB_D}\,,
\end{align}
where $\omega^\cA_D:=(\id\ot\cP_\cA)(|\omega_D\>\<\omega_D|)$ and  $\omega_D^\cB:=(\id\ot\cP_\cB)(|\omega_D\>\<\omega_D|)$. Here $I_c(A\>B):=S(B)-S(AB)$ is the coherent information~\cite{coherent}.

The correctable algebras have decompositions 
\begin{equation}
\cA\cong\bigoplus_{k}M_{n_k}(\mathbb{C})\ot I_{n_k'}\,,\quad \cB\cong\bigoplus_{l}M_{m_l}(\mathbb{C})\ot I_{m_l'},
\end{equation}
where $\sum_kn_kn_k'=\sum_lm_lm_l'=D$.  From these representations, it turns out that~\cite{takesaki}
\begin{equation}
\omega^\cA_D=\bigoplus_k p_k \frac{1}{n_k'}(I_{n_k})_{A_1}\ot (\omega_{n_k})_ {A_1A_2}\ot (\rho_{n_k'})_{A_2}
\end{equation}
with $p_k=\frac{n_kn_k'}{D}$ and some fixed states $\{\rho_{n_k}\}$. By simple calculations we have
\begin{align}
I_c(A_1\>A_2)_{\omega^\cA_D}&=S(A_2)_{\omega^\cA_D}-S(A_1A_2)_{\omega^\cA_D}\\
&=\sum_kp_k\log n_k-\sum_kp_k\log n_k'\\
&=\sum_kp_k\log \frac{n_k}{n_k'}\,.
\end{align}
Therefore, by applying the same argument for $\cB$, 
\begin{equation}\label{eq:dformula}
I(A_1:C_1|B_1)_{(1)}=\sum_kp_k\log \frac{n_k}{n_k'}+\sum_lq_l \log \frac{m_l}{m_l'}
\end{equation}
holds with $q_l=\frac{m_lm_l'}{D}$. One may obtain more detailed formula by employing $\cB'\subset\cA$.  The algebras satisfy $\cA=\cB'$ if and only if  $|\{k\}|=|\{l\}|$, $p_k=q_k$, $m_k=n_k'$ and $m_k'=n_k$. This leads $I(A_1:C_1|B_1)_{(1)}=0$. 

\subsection{Proof of Theorem~7}\label{ap:spt}
In this section, we prove Theorem~7. We start from deriving the group structure of tensor product logical unitaries. We then reveal a sufficient condition to imply the existence of SPT phase from the non-zero value of CMI. We finally show that all stabilizer states satisfy the sufficient condition. 

Recall that it is necessary to have a tensor product logical operator for $\phi^{(n)}$ to be in a SPT phase (under on-site symmetry). Let $\cG_\cA:=\cU(\cL_\cA)\cap\left(\cB(\cH_B)\ot\cB(\cH_C)\right)$ be the set of all tensor product logical operators. Since $\cL_\cA$ is the pre-image of $*$-homomorphism, it is a finite-dimensional  $C^*$-algebra and therefore $\cU(\cL_\cA)$ is a compact and connected Lie group.  $\cG_\cA$ is then a Lie subgroup of $\cU(\cL_\cA)$. 
\begin{prop} Define $\cC_\cA$ as the subgroup of logical operators
\begin{equation}\label{eq:identitycomp}
\cC_\cA=\left\{U_B\ot I_C, U_B\in\cL_\cA \right\}\,.
\end{equation}
$\cC_\cA$ is the identity component of $\cG_\cA$. 
\end{prop}
\noindent {\it Proof.} Suppose $U=e^{itO}\in\cU(\cL_\cA)$. Then $\ctE^\dagger(e^{itO})=e^{it\ctE^\dagger(O)}\in\cU(\cA)$ ($\ctE^\dagger$ is a $*-$homomorphism) and therefore $\ctE^\dagger(O)\in\cA$, i.e. $O\in\cL_\cA$. The Lie algebra of $\cU(\cL_\cA)$, which we denote by $\mathfrak{u}(\cL_\cA)$, is therefore Hermitian logical operator on $BC$. Furthermore, $\mathfrak{u}(\cL_\cA)$ should not contain operators like $I_B\ot O_C$ for $O_C\neq I_C$, since the input of $\ctE$ and $C$ is uncorrelated.  Therefore,
\begin{equation}
\mathfrak{u}(\cL_\cA)=\left\{L_B\ot I_C\,,\; L_{BC}\in\cL_\cA \left|  L_B=L_B^\dagger\,,\;L_{BC}=L_{BC}^\dagger \right.\right\}\,,
\end{equation}
where $\tr_BL_{BC}=\tr_CL_{BC}=0$. The Lie algebra of $\cG_\cA$ is a subalgebra of $\mathfrak{u}(\cL_\cA)$. For any non-local term $L_{BC}$, $e^{itL_{BC}}$ cannot be a tensor product for all $t\in\mathbb{R}$. Therefore, the identity component is given by $\{e^{itL_B\ot I_C}\}$, which is Eq.~(53). \qed

A well-known result on Lie group implies that the identity component is a closed normal subgroup and the  quotient  group  $\cG_\cA/\cC_\cA$ is a discrete group (see e.g. Ref.~\cite{EK} and references therein for more details). Since $\cG_\cA$ is compact, we obtain the following.
\begin{cor} $\cG_\cA/\cC_\cA$ is a finite group.
\end{cor}
The exactly same arguments holds for $\cG_\cB$ and $\cC_\cB$.  
We denote an abstract group isomorphic to $\cG_\cA/\cC_\cA$ by $G_1$ and similarly use $G_2$ for $\cG_\cB/\cC_\cB$. Under dual complementarity, we have $\cC_\cA=\cU(\cL_{\cB'})$.
Let ${\hat \cG}_\cA\leqslant\cU(\cA)$ be a subgroup of unitaries that have logical operators in $\cG_\cA$.  It is easy to check $\cU(\cB')\lhd{\hat \cG}_\cA$ and we obtain ${\hat \cG}_\cA/\cU(\cB')\cong G_1$ as well. 
For any $g\in G_1$ there exists $V(g)\neq I\in{\hat\cG}_\cA$ such that  the corresponding logical operator is in the form $U_B\ot U_C(g)$ with unitaries $U_B, U_C(g)\neq I_B, I_C$. $U_C(g)$ is independent of particular choice of element from the equivalence class $[V(g)]$, which has one-to-one correspondence with $g$ by definition. 

If $G_1$ represents the physical symmetry of the state, then $U_C(g)\in[V(g)]$ for any $g\in G_1$ by $C\cong A$. 
This guarantees that we obtain tensor product logical operators for arbitrary length $n$.  
However, in general there is no guarantee that $U_C(g)$ again has a tensor product logical operator. 
Actually, the non-trivial example shown in the main text violates this condition. 
We denote by $H_1$ the subgroup of $G_1$ such that $U_C(h)\in[V(h)]$ for any $h\in H_1$. 

A simple situation is that $H_1=G_1$. However, the group $G_1$ can still be small compare to $\cA$. To avoid this problem, we further assume ${\hat \cG}_\cA$ spans the whole correctable algebra. We show these two conditions are strong enough to obtain a non-trivial SPT phase from the value of CMI:
\begin{prop} \label{prop:sufficientspt}
Let $V$ be an isometry satisfying all of the following properties:
\begin{itemize}
\item dual complementarity
\item $H_1=G_1$, $H_2=G_2$
\item $Alg({\hat \cG}_\cA)=\cA$, $Alg({\hat \cG}_\cB)=\cB$
\end{itemize}
Then the MPS generated by $V$ is in a non-trivial $G_1\times G_2$ SPT phase after corse-graining a constant number of sites if, and only if, 
\begin{equation*}
I(A_1:C_n|B_n)_{(n)}>0,\; \forall n. 
\end{equation*}
\end{prop}
\noindent {\it Proof.} 
Since $H_1=G_1$, any $V(g)$ has a logical operator in the form $U_B\ot U_C(g)$ with $U_C(g)\in{\hat\cG}_\cA$. The equivalence class of unitaries $U_C(g)$ again consists a group isomorphic to $G_1$. It is clear that $f:g\mapsto [U_C(g)]$ is surjective from its definition. To show it is also injective, suppose that $U_C(g)=U_C(g')$ for $g,g'\in G_1$. Then we can find logical operators of $V(g)V(g')^\dagger$ in the form $U_B\ot I_C$, and thus $V(g)V(g')^\dagger\in\cU(\cB')$. This implies $[V(g)]=[V(g')]\Leftrightarrow g=g'$. Moreover, if  $U_C(g)=VU_C(g')$ for $V\in\cU(\cB')$, then $g=g'$. This is because we can apply $\cV_{C\to E'B'C'}$ and then logical operators of $U_C(g)$ and $VU_C(g')$ has the same unitary on $C'$, so the injectivity discussed before applies to this case as well. 
Therefore $f:g\mapsto [U_C(g)]$ is a well-defined group isomomorphism. 

Both $[V(g)]$ and $[U_C(g)]$ form ${\hat\cG}_\cA/\cU(\cB')$, but the labeling could be different, i.e. it might be true that $[V(g)]=[U_C(g')]$. 
The correspondence between $[V(g)]$ and $[U(g)]$ is a permutation (or automorphism) $Per:g'\mapsto g$ on $G_1$.  Therefore, there exists a constant $m\in\mathbb{N}$ such that $(Per)^m=id_{G_1}$.  By coarse-graining $m$ channels, we obtain the desired relation
\begin{equation}
\cV_{A_2\to B_mE_mC}(V(g)|\psi\>_{A_2})=(U_{B_m}(g)\ot V(g))\cV_{A_2\to B_mE_mC}(|\psi\>_{A_2})\,\; \forall g\in G_1
\end{equation}
with a unitary $U_{B_m}(g)$, where $B_m$ and $E_m$ are the coarse-grained systems. 
These arguments are also applicable for $G_2$, with possibly different coarse-graining scale $m'$. If $m\neq m'$, we coarse-grain ${\tilde m}:={\rm lcm}(m,m')$ sites. The unitaries $U_{B_{\tilde m}}(g_1)\ot U_{E_{\tilde m}}(g_2)$ for $(g_1,g_2)\in G_1\times G_2$, constracted in this way, form a unitary representation of $G_1\times G_2$. 

The states generated by $V$ satisfy $I(A_1:C_n|B_n)_{(n)}>0$ if and only if $\cB\subsetneq\cA'$ (Theorem~6) due to dual complementarity.  Moreover, $V(g_1)$ and $V(g_2)$ form a non-trivial  projective representation. By assumption, ${\hat \cG_{\cA}}$ and ${\hat \cG_{\cA}}$ contain the basis of $\cA$ and $\cB$. Moreover, 
for every $g_1,g_2$ $V(g_2)\in{\hat \cG_{\cB}}\subset\cB'$ and $V(g_1)\in{\hat \cG}_\cA\backslash\cU(\cB')$ and therefore there exists a pair $(g_1,g_2)$ such that $[V(g_1), V(g_2)]\neq0$ (note that the corresponding symmetery actions $U_{B_{\tilde m}}(g_1)\ot I_{E_{\tilde m}}$ and $I_{B_{\tilde m}}\ot U_{E_{\tilde m}}(g_2)$ commute each other). This completes the proof.  \qed

\subsubsection{Proof of Theorem~7}
\begin{proof} We show the theorem by proving that all the conditions in Proposition~\ref{prop:sufficientspt} are satisfied when $V$ is Clifford. 
Since $V$ is Clifford, $\cV_{A_2\to BEC}$ is an encoding map of a stabilizer code. In stabilizer codes, all the logical operators are spanned by logical Pauli operators~\footnote{A necessary and sufficient condition for an operator $O$ to be logical is $[O,S_i]=0$ for all stabilizer generators $S_i$. $O$ can be expanded in the product Pauli basis and it turns out that each non-zero part should commute with $S_i$ to satisfy the condition.}. The pre-image of $\ctE^\dagger$ is thus spanned by Pauli operators and it follows that $\cA=Alg\{P_i|\exists{\tilde P}_B\ot{\tilde P}_C\in\cL(P_i)\}$, where ${\tilde P}_B$ and ${\tilde P}_C$ are Pauli operators.  Let $D=2^K$ without loss of generality. $A$ is regarded as a $K$-qubit system and Pauli operators on $A$ are generated by $Z_i,X_i$ operators acting on $i$th qubit. 
Then, the generators of $\cA$ (up to a local Clifford) are summarized as a table:
\begin{align}
\cA\cong Alg\left(
\begin{array}{cccccccccccc}
1 & \cdots &l &l+1 &\cdots&m&m+1&\cdots&K \\
Z&\cdots&Z&Z&\cdots &Z&I&\cdots&I\\
I&\cdots&I&X&\cdots &X&I&\cdots&I\\
\end{array}
\right).
\end{align}
Here, the first column means $\cA$ contains $Z_1$, but not contains $X_1$. In the same way, $m+1$th column means $\cA$ does not contain both $Z_{m+1}$ and $X_{m+1}$. The commutant of $\cA$ is immediately given as 
\begin{align}
\cA'\cong Alg\left(
\begin{array}{cccccccccccc}
1 & \cdots &l &l+1 &\cdots&m&m+1&\cdots&K \\
Z&\cdots&Z&I&\cdots &I&Z&\cdots&Z\\
I&\cdots&I&I&\cdots &I&X&\cdots&X\\
\end{array}
\right).
\end{align}
From these expressions it is clear that $\cA=Alg({\hat \cG}_\cA)$. 
Let $g_{BC} (g_E)$ is the number of independent logical Pauli operators supported on $BC$ $(E)$. For stabilizer codes, it is known that they satisfy the formula $g_{BC}+g_{E}=2K$~\cite{Yoshida}. 
It is then clear that the number of logical operators found on $E$ is $l+2(K-m)$, which is the number of independent generators of $\cA'$. Since the correctable algebra corresponding to output on $E$ should be a subalgebra of $\cA'$, they must be equivalent. Therefore dual complementarity is satisfied. 

$H_1=G_1$ follows from $\cA=\cA_{\ctE}=\cA_{\ctE^{(2)}}$. Let us consider $P_i\in\cA\backslash\cB'$. For $n=1$, $P_i$ have a logical operator ${\tilde P}_{B_1}\ot{\tilde P}_{C_1}\in\cL(P_i)$ such that ${\tilde P}_C\neq I_C$. Since ${\tilde P}_C$ is also a Pauli operator, it has a logical Pauli operator on $B_2E_2C_2$. Suppose every such ${\tilde P}_C$ has no logical operator on $B_2C_2$. Then ${\tilde P}_C\in\cA_{\ctE}\backslash\cA'_{\ctF}$ but ${\tilde P}_C\notin\cA_{\ctE^{(2)}}\backslash \cA'_{\ctF^{(2)}}$, which conflicts to $\cA=\cA_{\ctE}=\cA_{\ctE^{(2)}}$ and $\cB=\cA_{\ctF}=\cA_{\ctF^{(2)}}$. Therefore $\cL({\tilde P}_C)$ contains operator on $B_2C_2$, which is also a Pauli operator and thus has a tensor product form. This proves $H_1=G_1$. The same arguments hold for $H_2$. 
\end{proof}

\appendix

\end{document}